\documentclass[preprint2]{aastex}
\usepackage{epsfig}
\slugcomment{accepted for publication in ApJL}

\shorttitle{A CORRELATION BETWEEN GALAXY LIGHT CONCENTRATION AND CLUSTER
DENSITY}

\shortauthors{Trujillo et al.}

\begin{document}

\title{A CORRELATION BETWEEN LIGHT CONCENTRATION AND CLUSTER LOCAL DENSITY
FOR ELLIPTICAL GALAXIES}

\author{{I. Trujillo}, {J.A.L. Aguerri}, {C.M. Guti\'errez}, {N. Caon} 
and {J. Cepa}}
\affil{Instituto de Astrof\'{\i}sica de Canarias, E-38200, La Laguna, 
Tenerife, Spain}
\email{itc@ll.iac.es, jalfonso@ll.iac.es, cgc@ll.iac.es, ncaon@ll.iac.es, 
jcn@ll.iac.es}

\begin{abstract} 

Using photometric and redshift data for the Virgo and  Coma clusters, we
present evidence for a correlation between the light concentration of 
elliptical galaxies (including dwarf ellipticals) and the local 3--D (i.e. 
non--projected) density of the clusters: more concentrated ellipticals are 
located in  denser regions.   The null hypothesis (i.e. the absence of any
relation) is rejected at a significance level better than  99.9\%. In order to
explain the observed relation, a power law  relating the galaxy light
concentration and the cluster 3--D density  is proposed.   We study how the
projection effects affect the form and dispersion of the  data--points in the
light concentration--projected density diagram. 
The agreement between our model and the observed data suggests that there is a 
paucity of dwarf elliptical galaxies in the cluster central regions.

\end{abstract}


\section{Introduction}

Analyzing the Abell Cluster 2443, Trujillo et al. (2001, hereafter T01) 
showed  evidence for a correlation between galaxy light concentration  and
local cluster surface density for elliptical galaxies: more centrally 
concentrated ellipticals appear to populate  denser regions. If this relation
is shown to hold for other clusters, it means that the qualitative morphology
density relation noted by previous authors (e.g. Dressler 1980; Dressler et al.
1997; Fasano et al. 2000) can be placed on a more quantitative basis, that is,
the detailed structure of the individual galaxies (beyond the broad
elliptical/spiral distinction) is related to their immediate 
environment/density.

That the structural properties of ellipticals are related to the properties of 
their parent clusters was  noted early on by Strom \& Strom (1978). The
characteristic sizes of  galaxies decrease by a factor of 1.5 in the denser
regions. This effect was explained by tidal disruption and high--speed impulse
encounters. However, as we discussed in T01, this mechanism does not  seem to
be the correct explanation for the correlation presented in our  previous
paper. Since mergers tend to increase the concentration of galaxies (White
1983; Barnes 1990, 1992) and bulges (Aguerri et al. 2001), we proposed 
this   mechanism as a possibly more viable explanation. 

By understanding how the properties of galaxies relate to those of their parent
clusters, we can hope to learn about the formation and evolution of both. In
this paper, the connection between  galaxy light concentration and 
cluster density is examined and confirmed for the Virgo and Coma clusters. We
pay special attention to the effects of projection on this relation, and show 
how its shape and its scatter can be easily explained.

\section{Galaxy data and measurements}

The data for the Coma cluster were taken from a quantitative morphological
analysis of the galaxies placed in the central region of this cluster, covering
an area of 0.28 square degrees (see details in Guti\'errez et al. 2002). The
images were obtained on 2000 April 25 and 27 using the Wide Field Camera at the
2.5 m Isaac Newton Telescope at the Observatorio del Roque de los Muchachos on
La Palma. The pixel scale is $0''.333$ pixel$^{-1}$, and the  seeing was
$1''.1$. The field was observed through R filter for a total  integration time
of 3900 s. B--R color information for the observed galaxies was obtained from
the Coma cluster galaxies catalog presented in Godwin, Metcalfe \& Peach
(1983). This catalog is complete down to $m_{B}=20$.

Briefly, the structural parameters of the elliptical galaxies were obtaining by
fitting a 2D  S\'ersic model $r^{1/n}$ (S\'ersic 1968) to the
observed galaxies. Both the ellipticity shapes of the galaxies and the effects
of seeing on the images were taken into account when fitting the model (details
of the parameter recovering method are  explained in T01). We used a Moffat
function with $\beta$=2.5 to describe the point--spread function. The
parameters were estimated with an error less than 10\% down to  R=17 
($m_{B}=19$), which we assume as our limiting magnitude for an accurate
morphological structure analysis. Assuming  $H_0=75$ km s$^{-1}$  Mpc$^{-1}$
(which we do throughout), and a redshift for Coma $z=0.023$, this implies an
absolute limiting magnitude $M_B=-15.84$ and an  observed field covering the
inner 500 kpc.

We then computed the concentration index of the best-fitting $r^{1/n}$ models
using a new index presented in Trujillo, Graham \& Caon (2001) and further 
developed in Graham, Trujillo \& Caon (2001).   This index measures the light
concentration within a profile's half--light  radius ($r_e$): it is the ratio
of flux inside some fraction $\alpha$ of the  half--light radius to the total
flux inside the half-light radius.  For an $r^{1/n}$ model, this index can be
analytically defined as \begin{equation}
C_{r_e}(\alpha)=\frac{\gamma(2n,b_n\alpha^{1/n})}{\gamma(2n,b_n)}, \label{TGC}
\end{equation} where $n$ is the shape parameter of the $r^{1/n}$ model and
$b_n$ is derived numerically from the expression $\Gamma (2n)=2\gamma(2n,b_n)$, 
with $\Gamma(a)$ and $\gamma(a,x)$  respectively the gamma function and the
incomplete gamma function (Abramowitz \& Stegun 1964).   The parameter $\alpha$
can be any value between 0 and 1, and defines what  level of concentration is
being measured. We used a value of $\alpha = 0.3$.

We computed the local cluster surface density around each galaxy in our sample
in the following way. Using the position information from the catalog   by
Godwin  et al. (1983), we compute the distance to the 10th nearest neighbor,
$r_{10}$, and derived the density as $\rho_{\rm proj} = 10/(\pi \times
r_{10}^2)$. In order to avoid contamination from field galaxies and to assure
uniform completeness on the whole area, we selected from that catalog only
those galaxies with $m_B \leq 20$ ($M_B < -14.84$), and satisfying the redshift
condition $4000 < cz < 10000$ km s$^{-1}$  (a 3$\sigma$ interval around the
mean cluster redshift). The redshift information was obtained from M. Colles
(private communication) from the data used in Edwards et al. (2002). For those
galaxies lacking velocity data, we use the color constraint $1 < B-R < 2$ (see
e.g. Mobasher et al. 2001).

The same calculations were carried out for the Virgo Cluster. The concentration
indexes were computed from the best-fitting Sersic $n$'s published in Caon,
Capaccioli \& D'Onofrio (1993), and in Binggeli \& Jerjen (1998); only
galaxies  classified as ellipticals or dwarf ellipticals brighter than
$m_B=15.45$ were  used. The projected densities were computed using the
galaxies listed in the Virgo  Cluster Catalog (Binggeli, Sandage \& Tammann
1985), from which we selected only those with confirmed membership and brighter
than $m_B=16.45$. The cut-off magnitudes used in Virgo correspond to  the
same limiting absolute  magnitudes used for Coma (assuming a Coma--Virgo
distance modulus of 3.50 mag from  D'Onofrio et al. 1997).

\section{A $C_{r_e}(1/3)$--$\rho$(3D) power law relation}

The relation between galaxy light concentration and local cluster surface
density is shown on Fig. 1. Solid points represent those elliptical galaxies
with $M_B<-17.5$ (22 from Virgo and 15 from Coma) whereas dwarf galaxies are
denoted by open points (21 from Virgo and 42 from Coma). Since $C_{r_e}(1/3)$,
as defined in equation 1, is a monotonic  function of the global shape
parameter $n$ (Trujillo et al.\ 2001a),  the local density -- $C_{r_e}(1/3)$
relation implies a relation  between the local density and $n$ as well. The
strength of these correlations are of course equal. We prefer to maintain the 
discussion in terms of the concentration parameter because it has a more 
tangible meaning that the index $n$.

\begin{figure}
\epsscale{0.5}

%\plotone{f1.eps}
\epsfig{file=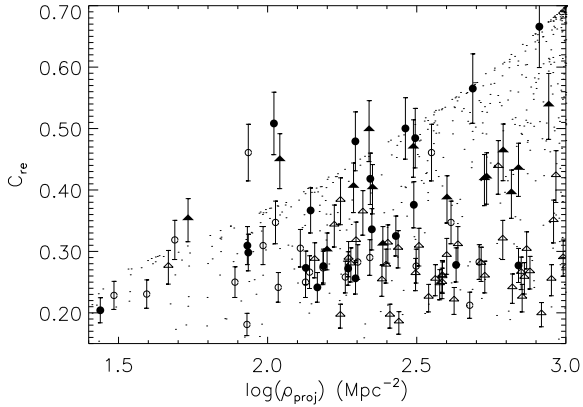, width=\columnwidth}

\caption[]{The light concentration index is plotted versus the
projected surface density. Solid points correspond to bright ellipticals
($M_B<-17.5$) in the Coma Cluster (triangles) and Virgo Cluster (circles). Open
points are dwarf galaxies in Coma (triangles) and Virgo (circles). Overplotted
on the observed points is a simulated cluster realization of 1000 points (small
dots) following Eq.~\ref{eq:dentgc}, with $\beta=1.3$ and $\delta=0.2$ (see
text  for details).}

\end{figure}

A point  we want to emphasize is the ``triangular'' form of the shape observed
between the galaxy light concentration and the cluster projected density. This
kind of relation is what one would expect in case a relation between  galaxy
light concentration and   cluster 3--D density is assumed.  To understand this,
we  note that projection effects will tend to mix both  more and less 
concentrated galaxies at higher projected densities, but at  lower projected
densities only galaxies with low concentrations will be seen,  i.e. no
high-concentration objects will appear in low density environments.

 To illustrate this, we
study what the shape and the scatter in the correlation look like by 
constructing 3--D simulated spherical clusters. For simplicity, we use the 
generalized King model density profile for the 3--D galaxy distribution:
\begin{equation}
\frac{\rho(r)}{\rho(r_c)}=\frac{2^{\beta+1/2}}{(1+(r/r_c)^2)^{\beta+1/2}}
\label{eq:density}
\end{equation}

where $r_c$ is the core radius and $\beta$ is a parameter that models the 
``tail''  of the profile. The extension of the tail decreases with 
increasing $\beta$. This model has a flat (core) behavior at $r<r_c$.

Each galaxy in the realization has associated a galaxy light concentration as a
function of the cluster 3--D density as follows:

\begin{equation}
C_{r_e}(\rho(r)/\rho(r_c))=C_{r_e,max}\bigg(\frac{1}{M}
\frac{\rho(r)}{\rho(r_c)}\bigg)^{\delta}
\label{eq:dentgc}
\end{equation}
where $C_{r_e,max}$ and $M$ are the maximum values of the concentration and the
cluster density, respectively.
The maximum value of $C_{r_e}$ is reached at the center $r=0$. Based on the
highest values of the concentration that we observe in our measurements,
we set $C_{r_e,max}=0.7$. On the other hand, M=2$^{\beta+1/2}$ for the King 
model.

In our models we assume that the distribution of the  artificial galaxies which
are used to evaluate the local density (i.e. the deeper sample) is described
also by the King model.  To calibrate the density of  this deep sample we have
imposed to the model that $\log$$\rho_{\rm proj}$(0)=3 (where $\rho_{proj}$(0) is 
the projected density at the center of the cluster). This  value is close to the
highest density  that we observe in our observational data (see Fig. 1).

Following Eq. (3), each  point ($\log (\rho_{\rm proj})$, $C_{r_e}$) in Fig. 1 can
be understood  (one--to--one) in terms of the pair ($R,r$), where $r$ is the
3--D radial distance  and $R$ is the  projected 2--D radial distance associated
to this point in the cluster model.

\begin{figure}
\epsscale{0.5}

%\plotone{f2.eps}
\epsfig{file=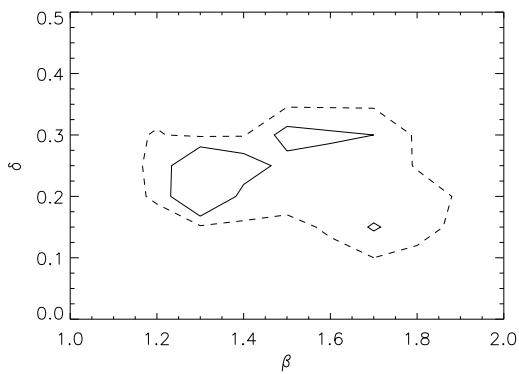,angle=0, width=\columnwidth}

\caption[]{The contours of confidence level 68\% (solid line) and 95\% (dashed
line) associated to the proposed model are shown.}

\end{figure}

The comparison between the simulated model point distribution and
the observed point distribution is done by using the generalization of the
Kolmogorov--Smirnov (K--S) test to two--dimensional distributions 
(Fasano \& Franceschini 1987).
The two--dimensional K--S statistic used (D) is defined as the maximum
difference (ranging both over data points and over quadrants) of the
corresponding fractions between the data and the model.
In order to match the observed point distribution intervals (see Fig. 1), we
restrict the above comparison  to the $0.18<C_{r_e}<0.70$ and
$1.4<\log\rho_{\rm proj}<3$ intervals.

The cluster simulations -- described in the Appendix -- are repeated 1000 times, 
so as to build a distribution function {\sl DF} for the quantity $D_{\rm max}$
(the largest difference of D for each cluster simulation). Finally, we repeat
the  process, this time using our observed data-points,  to obtain $D_{\rm
max,obs}$.

Following standard statistical methods, we evaluate the probability associated 
with the measured value of $D_{\rm max,obs}$ by calculating the fractional 
area under the {\sl DF} curve where the frequency is equal to or less than 
that of $D_{\rm max,obs}$. 

On Fig. 2 we plot the isocontours of confidence level 68\% and 95\% associated
to our models. To illustrate how well one of the ``acceptable'' models can 
reproduce the shape of the data distribution, a realization of the model 
$\beta=1.3$ and $\delta=0.2$ (using 1000 points) is over--plotted on the observed 
distribution in the $\log \rho_{\rm proj} - C_{re}$ diagram in Fig 1.  
It  clearly shows that the shape and the dispersion present in Fig. 1 is just  
the product of the projection effects, even when starting from perfectly  
noiseless 3--D relations, such as the one in Eq.~\ref{eq:dentgc}. 

By looking at Fig. 1, the toy  galaxies crowd near the diagonal (in particular the
upper right corner),  while most of the real galaxies (in particular dEs) are
located in the bottom  part of the plot. To explain this different distribution
we must note first that when  constructing the artificial clusters we ignore
the actual luminosity function of Es and dEs in the cluster
(i.e. artificial galaxies are placed in the clusters just taking into account the
density profile). 

It is clear that any improved version of the present models (out of the scope
of this paper) would have to take into account the fact that the dE galaxies are 
more abundant than the elliptical galaxies.
Moreover, as the dE galaxies are also less concentrated objects than the E
galaxies (see e.g. the luminosity-concentration diagram in Fig. 8 of Graham et 
al. 2001), one can expect that in observed galaxy distributions as the one 
presented in Fig. 1 most of the galaxies are found in the bottom  part of the 
plot.

The distribution of artificial galaxies  is also a function of the assumed 
symmetry used to describe the cluster (in our case spherical symmetry). 
The diagonal line in Fig. 1 is populated by the artificial galaxies which lie 
over the perpendicular plane to the line of sight passing through the cluster 
center. 
The fact that the structure of real clusters departs significantly from
the spherical simmetry of our simple model accounts for the differences between
the observed and simulated distributions.

To evaluate the null hypothesis (i.e. the absence of any relation between the
galaxy light concentration and the cluster density) we have constructed 3--D
clusters where each galaxy has a concentration index independent of the density
and randomly distributed between $0<C_{r_e}<0.7$. The null hypothesis is 
rejected at a confidence level $>$99.99\%. This is the main result of this 
paper: whatever the exact form of the relation between galaxy light 
concentration and cluster density may be, it is clear that such a  relation 
exists. Also, we have checked the consistency of our data by making an internal
comparison (based on a 2-D K-S test) between the distribution of Virgo and
Coma  galaxies.  The hypothesis that the two samples come from the same parent
population turns out to be acceptable (rejection probability  $\sim$0.3).

\section{Discussion}

A  discussion of the mechanisms that may be acting to generate the above
correlation was presented in T01. There, we explained that a mechanism based on
mergers is favored over other scenarios based on tidal friction and
high--speed impulse encounters.  In any case, in the center of galaxy clusters
different mechanisms  may be operating to reshape the form of the galaxies, and
no single mechanism is expected to account for the whole range of observed
properties.

From the model that we have proposed to analyse the distribution of galaxies in
Fig. 1, it follows that dE galaxies do not populate the central region of the
clusters. Using Eqs.~\ref{eq:density} and \ref{eq:dentgc} it is easy to
determine the radial distance to the center of the cluster once the galaxy
concentration  is known. Assuming that the maximum galaxy light concentration of
dEs is 0.4 (see. Fig. 1) and that the acceptable models are in the range
$1.2<\beta<1.8$ and $0.15<\delta<0.35$ (see Fig. 2), dE galaxies are
removed from the center of the clusters out to a radius $\sim 1-3r _c$.

Other authors (Secker, Harris \& Plummer 1997; Gregg \& West 1998; Adami et
al. 2001; Andreon 2002) have found independent evidence which supports this
result. This would be an indication that in the  denser cluster
environments the dEs (which are only weakly gravitationally bound) can be
destroyed by tidal disruption.
For the King model, we can make a direct comparison between our results and the
cluster evolutionary model proposed by Merritt (1984), in which tidal forces in
the cluster center disrupt the galaxies. The minimum size that a galaxy
must have in order not to be tidally disrupted is $\sim 15 h^{-1}$ kpc.
The maximum of the tidal disruption forces is reached at $r\sim r_c$. Thus,
this scenario predicts that dEs are easily destroyed, while the larger
and more massive elliptical galaxies are able to survive in the 
center of the cluster. 
Our results are in good agreement with this.

Another important result which follows from the kind of law suggested in 
Eq.~\ref{eq:dentgc} is the possibility of making a 3--D reconstruction of the 
elliptical galaxies in the cluster. 
If the spherical assumption for the density profile of the cluster
is good enough, then by measuring the galaxy light concentration of each
elliptical it will be possible to determine its 3--D radial distance. 
Of course, this kind of reconstruction is {\it a priori} possible for any law 
which relates a galaxy property (e.g. concentration, size, color, etc.) with 
the cluster environment density.

\acknowledgments

We thank Helmut Jerjen for providing us with an updated electronic copy of 
the Virgo Cluster Catalog,  Matthew Colless who kindly sent us his catalog  of
Coma Cluster galaxies in advance of publication, and Antonio Mar\'{\i}n who
amiably provided us with the observations of the Coma cluster core. We also
thank  Juan Betancort and Alister W. Graham for their useful comments.
We are specially grateful to Peter Erwin who proofread versions of
this manuscript. Finally, the authors are grateful to the anonymous referee for
the valuable comments that helped us to improve the rigor and clarity of this
paper.

\appendix

\section{The cluster simulations}

We carried out cluster simulations as follows.
\begin{enumerate}
\item We populate each cluster realization with the same number of artificial
galaxies as the observed ones (99), having radial distances in accord with the
density law in Eq.~\ref{eq:density}, and assign to them the concentration index
given by Eq.~\ref{eq:dentgc}.
The data-points are then projected on the $\log\rho_{\rm proj}-C_{re}$ diagram.
\item For each point $(x_i,y_i)$ in the diagram\footnote{In our case
($\log \rho_{\rm proj}(i)$, $C_{r_e}(i)$).}, we compute the fractional number of
data-points in the four quadrants $f_a(x>x_i, y>y_i)$, $f_b(x<x_i, y>y_i)$,
$f_c(x<x_i, y<y_i)$ and $f_d(x>x_i, y<y_i)$.\footnote{By construction,
$f_a+f_b+f_c+f_d=1$.}.
\item We then compute, for the same points, the expected fractions given by 
our density model, which can be determined analytically using the expressions:

\begin{equation}
f_a=F(r^*)-\frac{\int_{R^*}^{r^*}\cos(\arcsin(R^*/u))\rho(u)u^2du}
{\int_{0}^{\infty}\rho(u)u^2du}
\end{equation}
\begin{equation}
f_a+f_b=F(r^*)
\end{equation}
\begin{equation}
f_c+f_d=1-F(r^*)
\end{equation}
\begin{equation}
f_d=1-F(r^*)-\frac{\int_{r^*}^{\infty}\cos(\arcsin(R^*/u))\rho(u)u^2du}
{\int_{0}^{\infty}\rho(u)u^2du}
\end{equation}
where ($R^*,\;r^*$) are the values of projected  and the 3--D radial  distances
respectively associated to the observed point ($\log\rho_{\rm proj}$(i), C$_{r_e}$(i)),
and $F(r)$ is the cumulative distribution function defined 
(for a radial symmetric model) as: 
\begin{equation}
F(r)=\frac{\int_{0}^{r}\rho(u)u^2du}
{\int_{0}^{\infty}\rho(u)u^2du}
\end{equation}
with $\rho(u)$ given by Eq~\ref{eq:density}.
For the King model proposed above, $F(r)$ can be derived analytically when
$1<\beta<2$:

\begin{equation}
F(r)=\frac{4}{3\sqrt{\pi}}\bigg(
\frac{r}{r_c}\bigg)^{3}
\frac{\Gamma(1/2+\beta)_2F_1(3/2,\beta+1/2,5/2,-(r/r_c)^2)}
{\Gamma(\beta-1)}
\end{equation}

where $_2F_1(a,b;c,z)$  is the hypergeometric function (Abramowitz \& Stegun
1964, p. 556).

\item For each point $(x_i,y_i$) we find the maximum difference between the 
observed and expected fractions in the four quadrants. 
This is repeated for all data-points, so as to obtain the largest difference 
$D_{\rm max}$ for each cluster simulation.
\end{enumerate}


\begin{thebibliography}{}
\bibitem[Abramowitz (1964)]{abra64} Abramowitz M., Stegun I., 1964, Handbook of 
   Mathematical Functions. Dover, New York
\bibitem[Adami(2001)]{ada01} Adami, C., Mazure, A., Ulmer, M.P. \& Savine, C.,
   2001, A\&A, 371, 11
\bibitem[Aguerri(2001)]{ague01} Aguerri, J. A. L., Balcells, M. \& Peletier, R.
   F., 2001, A\&A, 367, 428
\bibitem[Andreon(2002)]{and02} Andreon, S., 2002, A\&A, 382, 821
\bibitem[Barnes(1990)]{bar90} Barnes, J. 1990, in Dynamics and Interactions of Galaxies, ed. R. Wielen (Heidelberg:
   Springer--Verlag), 232
\bibitem[Barnes(1992)]{bar92} Barnes, J. 1992, \apj, 393, 484
\bibitem[Binggeli(1985)]{bin85} Binggeli, B., Sandage, A., Tammann, G. A., 1985,
   AJ, 90, 1681.
\bibitem[Binggeli \& Jerjen(1998)]{1998A&A...333...17B} Binggeli, B.~\& 
   Jerjen, H.\ 1998, \aap, 333, 17. 
\bibitem[Caon et al (1993)]{Cao93} Caon, N., Capaccioli, M., \& D'Onofrio, M.
   1993, \mnras, 265, 1013
\bibitem[D'Onofrio et al (1993)]{Don93} D'Onofrio, M., Capaccioli, M., 
   Zaggia, S.R.,  \& Caon, N. 1997, \mnras, 289, 847
\bibitem[Dressler (1980)]{dre80} Dressler, A. 1980, \apj, 236, 351
\bibitem[Dressler et al (1997)]{dre97} Dressler, A.,  Oemler, A. Jr., Couch, W.
   J., Smail, I.,Ellis, R. S., Barger, A., Butcher, H., Poggianti, B. M.,
   Sharples, R. M. 1997 \apj, 490, 577
\bibitem[Edwards (2002)]{edw02} Edwards, S. A., Colless, M.,Bridges, T J., 
Carter, D., Mobasher, B., Poggianti, B. M., 2002, \apj,567, 178.
\bibitem[Fasano (1987)]{fa87} Fasano, G., \& Franceschini, A., 1987, \mnras,
   225, 155
\bibitem[Fasano (2000)]{fa00} Fasano, G., Poggianti, B. M., Couch, W.J.,
   Bettoni, D., KJaergaard, P., \& Moles, M. 2000, \apj , 542, 673
\bibitem[Godwin et al. (1983)]{god83} Godwin, J. G., Metcalfe, N., 
 Peach, J. V., 1983, \mnras, 202, 113.
\bibitem[Graham (2001)]{gra01b} Graham, A. W., Trujillo, I., \& Caon, N., 2001,
   \aj, 122, 1707
\bibitem[Gregg (1998)]{greg98} Gregg, M. D., West, M. J. 1998, Nature, 396, 549
\bibitem[Gutierrez (2002)]{gut02} Guti\'errez, C.M., Trujillo, I., Aguerri,
   J.A.L.,  Colless, M., Caon, N., Cepa, J., 2002, in preparation
\bibitem[Merrit(1984)]{mer84} Merritt, D., 1984, \apj, 276, 26
\bibitem[Mob (2001)]{mob01} Mobasher, B. et al, 2001, \apjs, 137, 279
\bibitem[Secker (1997)]{sec97} Secker, J., Harris, W. E., Plummer, J. D., 1997,
   PASP, 109, 1377
\bibitem[S\'ersic(1968)]{ser68}  S\'ersic, J.  1968, Atlas de Galaxias 
   Australes C\'ordoba: Obs. Astron\'omico
\bibitem[Strom (1978)]{str78} Strom, S. E. \& Strom, K.M. 1978, \apj, 225, L93
\bibitem[Trujillo et al.(2001b)]{tru01a} Trujillo, I., Graham A.W. \& Caon, N.,
   2001a, \mnras, 326, 869.
\bibitem[Trujillo et al.(2001b)]{tru01b} Trujillo, I., Aguerri, J. A. L., 
   Guti\'errez, C. M. \& Cepa, J. 2001, \aj, 122, 38 (T01)
\bibitem[white(1983)]{whi83} White, S.D.M. 1983, in Dynamics and Interactions 
   of Galaxies, ed. E. Athanassoula (Dordrecht; Reidel), 337

\end{thebibliography}
\end{document}